\documentclass[journal=jacsat,manuscript=article,layout=twocolumn]{achemso}

\usepackage{geometry}

\setkeys{acs}{articletitle=true}
\usepackage{lettrine}

\usepackage[version=3]{mhchem} % Formula subscripts using \ce{}
\usepackage[T1]{fontenc}       % Use modern font encodings

\usepackage[fontsize=10pt]{scrextend}

\usepackage{newtxtext}
\usepackage{bm} 
\usepackage[labelfont=bf]{caption} % bold ``Figure''

\usepackage[hyperfootnotes=false]{hyperref}	 % make refs, cites, links in color
\hypersetup{
  colorlinks,
  citecolor=blue,
  linkcolor=blue,
  urlcolor=blue
}

\usepackage{titlesec} % inline subsections
\titleformat{\subsection}[runin]
{\normalfont\bfseries}{\thesubsection{.}}{1em}{}[.]

\titleformat{\section}
{\normalfont\sffamily\large\bfseries}{\thesection{.}}{1em}{\MakeUppercase}

\usepackage[table]{xcolor}
\usepackage{graphicx}
\usepackage{amsmath,fullpage,amssymb}
\usepackage{graphics,epsfig}
\usepackage{makeidx}
\usepackage{gensymb}
\usepackage{siunitx}
\usepackage{subcaption}

\usepackage{soul}	% for strikethrough font

\usepackage{lineno}     %  for line lumbers

\def\ln{{\operatorname{ln}}}

\def\o{\mathrm{o}}
\def\ow{\mathrm{ow}}
\def\w{\mathrm{w}}
\def\cav{\mathrm{cav}}

\def\Eq{eq}
\def\Eqs{eqs}

\def\Fig{\textcolor{blue}{Figure}}

\newcommand{\kB}{k_\textrm{B}}

\newcommand{\del}[1]{\textcolor{orange} {}}
 
\newcommand{\trm}[1]{{\textrm{#1}}}

\newcommand{\comment}[1]{\textcolor[rgb]{0,0.,0}{#1}}

\SectionNumbersOn
\let\oldmaketitle\maketitle
\let\maketitle\relax

\title{\flushleft 
Impact of Nanoscopic Impurity Aggregates on Cavitation in Water
}

\author{Marin \v{S}ako}
\affiliation{\rm\small Jo\v{z}ef Stefan Institute, Jamova 39, 1000 Ljubljana, Slovenia}
\alsoaffiliation{\rm\small Faculty of Mathematics and Physics, University of Ljubljana, Jadranska 19, 1000 Ljubljana, Slovenia}

\author{Roland R.\ Netz}
\affiliation{\rm\small Fachbereich Physik, Freie Universit\"{a}t Berlin, Berlin 14195, Germany }

\author{Matej Kandu\v{c}}
\affiliation{\rm\small Jo\v{z}ef Stefan Institute, Jamova 39, 1000 Ljubljana, Slovenia}
\email{matej.kanduc@ijs.si}

\begin{document}
\pagenumbering{arabic}
\noindent

\parindent=0cm
\setlength\arraycolsep{2pt}

\twocolumn[	% make wide abstract
\begin{@twocolumnfalse}
\oldmaketitle

\begin{abstract}\small
The stability of water against cavitation under negative pressures is a phenomenon known for considerable discrepancies between theoretical predictions and experimental observations. Using a combination of molecular dynamics simulations and classical nucleation theory, we explore how nanoscopic hydrocarbon droplets influence cavitation in water. Our findings reveal that while a macroscopic volume of absolutely pure water withstands up to $-$120 MPa of tension, introducing a single nanoscopic oil droplet, merely a few nanometers in radius, brings this cavitation threshold to around $-$30 MPa, closely matching the values typically observed in highly controlled experiments. The unavoidable presence of nanoscopic hydrophobic impurities, even in highly purified water used in experiments, imposes a practical limit on achieving the theoretical tensile strength in realistic settings. More broadly, our study highlights the profound impact of nonpolar residues on nucleation phenomena and enhances our understanding of metastability in real-world systems. \del{These insights are significant for a wide range of scientific and industrial applications, from plant physiology to advanced material processing.}

\bf{KEYWORDS:} \sl{cavitation, bubble, droplet, water, oil, molecular dynamics simulation}
\vspace{5ex}
\end{abstract}
\end{@twocolumnfalse}]

\maketitle
\setlength\arraycolsep{2pt}
\small

%\linenumbers            %  enable line lumbers

\section{Introduction}
Liquids can be brought into a metastable state relative to their vapor in two ways: either by superheating them beyond their boiling temperature or by stretching them below their saturated vapor pressure. Eventually, they will revert to equilibrium through the nucleation of vapor bubbles, a phenomenon known as cavitation. The ability of a liquid to withstand mechanical tension and maintain negative pressures highlights the cohesive forces among its molecules. Water is a paradigmatic example, owing to its exceptional cohesion due to hydrogen bonds~\cite{caupin2006cavitation}.

The phenomenon of negative pressures in liquids may seem counterintuitive at first, yet it commonly occurs in nature. A famous example is the sap in plants, reaching negative pressures as deep as $-$10 MPa~\cite{stroock2014physicochemical}, which helps it to ascent against gravity.
%Plants, as well as pathogens residing in their xylem, must avoid cavitation to survive.
Octopuses and squids utilize negative pressures in their suckers for effective adhesion~\cite{smith1991negative}, whereas ferns exploit negative pressure for spore dispersal through a catapult-like mechanism~\cite{noblin2012fern}.
In porous media, soil, and microfluidic capillaries, water can experience tension because of curved air--water interfaces or evaporation~\cite{wheeler2008transpiration, pagay2014microtensiometer, vincent2014drying}.
% errosion
When cavitation bubbles form under tension, they collapse violently when the tension is released, generating microjets and shock waves, which can damage nearby surfaces, causing erosion in turbomachinery or harm biological tissues~\cite{dular2004relationship, reuter2022cavitation, adhikari2015mechanism, peters2015numerical}.
Conversely, cavitation is the basis of lithotripsy, a medical procedure that uses acoustic shock waves to break down kidney stones and other calcifications within the body, the exact mechanism of which is still debated~\cite{coleman1989survey, caupin2006cavitation}.

% research
The study of water under tension and cavitation, dating back to the 17th century~\cite{caupin2006cavitation}, has inspired researchers to understand not only the metastable phase diagram of water and its thermodynamic anomalies, but also shed light on the nature of first-order transitions as a fundamental phenomenon~\cite{caupin2015escaping}.
Cavitation inception, a key process in this scope, is the stochastic nucleation of a vapor bubble. The interplay between the volume and surface energies of the bubble creates an energy barrier that has to be surmounted by the thermal fluctuations within the system.
This concept forms the cornerstone of Classical Nucleation Theory (CNT)~\cite{debenedetti1996metastable}, which suggests that water at room temperature should sustain negative pressures between about $-$120 and $-$160 MPa for extended periods of time~\cite{fisher1948fracture, caupin2005liquid, caupin2006cavitation, caupin2013stability, azouzi2013coherent}.
\comment{This remarkably high degree of metastability has been confirmed also by recent computer simulation studies~\cite{menzl2016molecular, wang2017cavitation, xie2022study, p2023water}.}

However, in less controlled experiments, cavitation in water occurs near saturated vapor pressure (around 2~kPa), preventing it from reaching a significantly metastable state.
The prevalent understanding is that cavitation initiates at the weak points within a liquid, such as hydrophobic surfaces, impurities, and especially pre-existing bubbles~\cite{harvey1944bubble, atchley1989crevice, jones1999bubble}, \comment{as demonstrated by systematic experiments~\cite{borkent2009nucleation} and  simulations~\cite{gao2021effects}.}
Indeed, purifying and degassing water enhances its sustainability against tension and  allows it to withstand considerable negative pressures.
However, even the most rigorous purification methods combined with acoustic cavitation or shock waves techniques that avoid solid boundaries seem to consistently reach a threshold of only about $-$30 MPa~\cite{caupin2006cavitation, caupin2013stability, caupin2015escaping, LAUTERBORN2023}, which is by a factor of 4--5 below theoretically predicted values.
A notable exception involves microsized inclusions in minerals, similar to the Berthelot tube method~\cite{zheng1991liquids, alvarenga1993elastic, azouzi2013coherent}, where negative pressures up to $-$140 MPa can be achieved, though the reasons for this are not fully understood.

A prevailing hypothesis attributes the gap between theory and experiment to nanoscale nuclei suspended even in highly purified water, as achieving absolute purity in large liquid volumes is practically impossible~\cite{roger2012hydrophobic, maali2017viscoelastic, gao2021effects, zhang2021critical}. \comment{Indeed, recent MD simulations indicate that solid particles can promote cavitation within simulation volumes by disrupting the hydrogen-bond network~\cite{li2019cavitation}. }
Nevertheless, it remains uncertain whether such trace contaminants meaningfully contribute to the cavitation of macroscopic systems, given that their minute amounts may have a limited influence on the entire system.
While testing this ``contamination hypothesis'' via experiment is \comment{very challenging, theoretical approaches offer a vital alternative.}

\comment{In this study, we quantify the influence of nanoscale impurities, specifically hydrophobic oil droplets, on the cavitation pressure of water. Using MD simulations combined with CNT, we demonstrate that a single nanoscopic oil droplet, only a few nanometers in radius, suspended in a large volume of water brings the cavitation threshold to around $-$30 MPa, closely matching experimental observations. These results represent the first quantitative perspective into how non-soluble nanoscale impurities set the reported practical limit on water’s tensile strength in realistic environments, thereby helping to bridge the gap between theory and experiment.
Our simulations provide insights into cavitation inception at hydrophobic droplets, relevant to applications such as ultrasonic emulsification. They clarify how hydrophobic inclusions influence cavitation thresholds in water, offer an explanation for why the theoretical limit is reached only in microscale systems, and shed light on mechanisms of cavitation resilience in plant sap.}

\section{Results and discussions}

\subsection{Homogeneous cavitation in absolutely pure water}
CNT is foundational in the study of homogeneous cavitation, offering a framework to understand the transition from a metastable to a stable phase in liquids~\cite{caupin2006cavitation, azouzi2013coherent, menzl2016molecular}. 
CNT posits that cavitation inception occurs with the stochastic formation of a  bubble within the bulk liquid, which occurs under substantial negative pressures.
The free energy of an expanding spherical cavity of radius $r$ under negative pressure $p$ is expressed as
\begin{equation}
    \label{eq:G_tolman}
    G_\trm{w} (r) = 4 \pi   \gamma(r) r^2 + \frac 43 \pi p r^3
\end{equation}
where the first term is associated with the free energy of creating the bubble interface and the second term to the work gained by performing the volume expansion against the ambient pressure.
Achieving accurate quantitative agreement with molecular simulations requires considering the impact of interface curvature on surface tension, $\gamma(r)$, as highlighted before~\cite{menzl2016molecular, Kanduc10733}. We describe the curvature-dependent surface tension as $ \gamma(r) = \gamma_{\w} (1 + 2\delta_\w/r)$, where $\gamma_{\w}$ is the surface tension of the flat water--vapor interface and $\delta_\w$ the Tolman length. 
\comment{
The Tolman length quantifies the deviation in surface tension between a curved and a planar interface, which becomes significant for nanometer-scale radii.} This expression represents the first-order Taylor expansion of Tolman's original expression~\cite{tolman1949effect}, which is applicable in scenarios of small curvatures ($r\gg|\delta_\w|$). Equation~\ref{eq:G_tolman} neglects the effect of vapor inside the cavity, which is reasonable given that the typical cavitation pressures are much higher in magnitude (tens of MPa) than the saturated vapor pressure at room temperature (several kPa).

The competition between the two terms in \Eq~\ref{eq:G_tolman}, each of a different sign (note that $p<0$) and scaling differently with the bubble radius, results in a free energy maximum $G_\w^* $ at the critical radius $r^*$. 
To first-order in $\delta_\w$, the critical radius, determined by the condition $\mathrm{d}G_\w / \mathrm{d} r = 0$, expresses as
\begin{equation}
    \label{eq:r_star_tolam}
    r^* = -\frac{ 2\gamma_{\w}}{p} +\delta_\w
\end{equation}
and the free energy barrier as
\begin{equation}
    \label{eq:dG_star_approx}
     G^*_\w = \frac{16 \pi \gamma_{\w}^3}{3p^2} - \frac{16\pi \gamma_{\w}^2\delta_\w}{p}
\end{equation}
A bubble exceeding the radius $r^*$ will spontaneously grow, leading to a spread over the entire system.

Cavitation is a thermally activated stochastic process in which a bubble in a given volume spontaneously reaches the critical size.
According to reaction rate theory, the rate of cavitation within this volume
is  given by~\cite{hanggi1990reaction, Kanduc10733}
\begin{equation}
    \label{eq:k}
    k_\w = \kappa_\w V_\w\,e^{-\beta G^*_\w}
\end{equation}
Here, the kinetic prefactor $\kappa_\w V_\w$ denotes the frequency of cavitation attempts, which is proportional to the volume of the liquid $V_\w$. The quantity $\kappa_\w$ is the attempt frequency density, an intensive property of the liquid.
The Boltzmann factor, $\exp(-\beta G^*_\w)$, accounts for the probability that an individual attempt reaches the critical size.

The survival probability that the system has not yet cavitated decays exponentially with time as $\exp(-k_\trm{w}t)$. Thus, the cavitation rate $k_\trm{w}$ represents the inverse of the mean time for cavitation to occur, $\tau=k_\trm{w}^{-1}$. 
In practical terms, it is often more useful to discuss cavitation in the context of cavitation pressures rather than cavitation times. Specifically, we define the cavitation pressure $p_\cav$ as the pressure at which the likelihood of cavitation reaches $e^{-1}$ for a given time $\tau$. Note that an alternative definition of the cavitation pressure, at which the probability reaches $1/2$ is also used in the literature~\cite{caupin2006cavitation}, leading to a negligible difference in cavitation pressure.
From the expressions for the free energy barrier (\Eq~\ref{eq:dG_star_approx}) and cavitation rate (\Eq~\ref{eq:k}) and considering the result for the mean cavitation time $\tau = k_\trm{w}^{-1}$, we can compute the cavitation pressure.
If we disregard the curvature dependence of surface tension (i.e., setting $\delta_\w = 0$), as traditionally done in the literature, we get the following, well-established CNT prediction for the cavitation pressure~\cite{caupin2006cavitation, herbert2006cavitation}
\begin{equation}
    \label{eq:tens_strength_old}
    {p}^{(0)}_\cav  = - \sqrt{\frac{16 \pi  \gamma_\w^3}{3 \kB T\,  \ln (\kappa_\w V_\w \tau)}}
\end{equation}
where $\tau$ is considered as the observation time.
Incorporating the first-order curvature correction, results in the following modified expression for the cavitation pressure 
\begin{equation}
    \label{eq:p_cav_vs_t}
    p_\cav  = {p}^{(0)}_\cav\left( 1 - \frac{3 \delta_\w}{2\gamma_\w} {p}^{(0)}_\cav\right)
\end{equation}
This modified expression is accurate when the second term in parentheses is much less than 1, a condition met in our system.

Determining the attempt frequency density, $\kappa_\trm{w}$, and the Tolman length, $\delta_\trm{w}$, presents a considerable challenge, as they are not amenable to simple continuum descriptions. Instead, these parameters require molecular-level considerations for an accurate estimation.
While various theoretical estimates exist for the attempt frequency~\cite{blander1975bubble, pettersen1994experimental}, their reliability often remains questionable.
Fortunately, advancements in molecular simulations provide a robust solution to this challenge, offering detailed microscopic insights into cavitation and enabling precise quantification of relevant physical quantities~\cite{Kanduc10733, menzl2016molecular, allen2009forward}.

\begin{figure}[h!]
    \begin{subfigure}[b]{0.45\textwidth}
    \includegraphics[width=\textwidth]{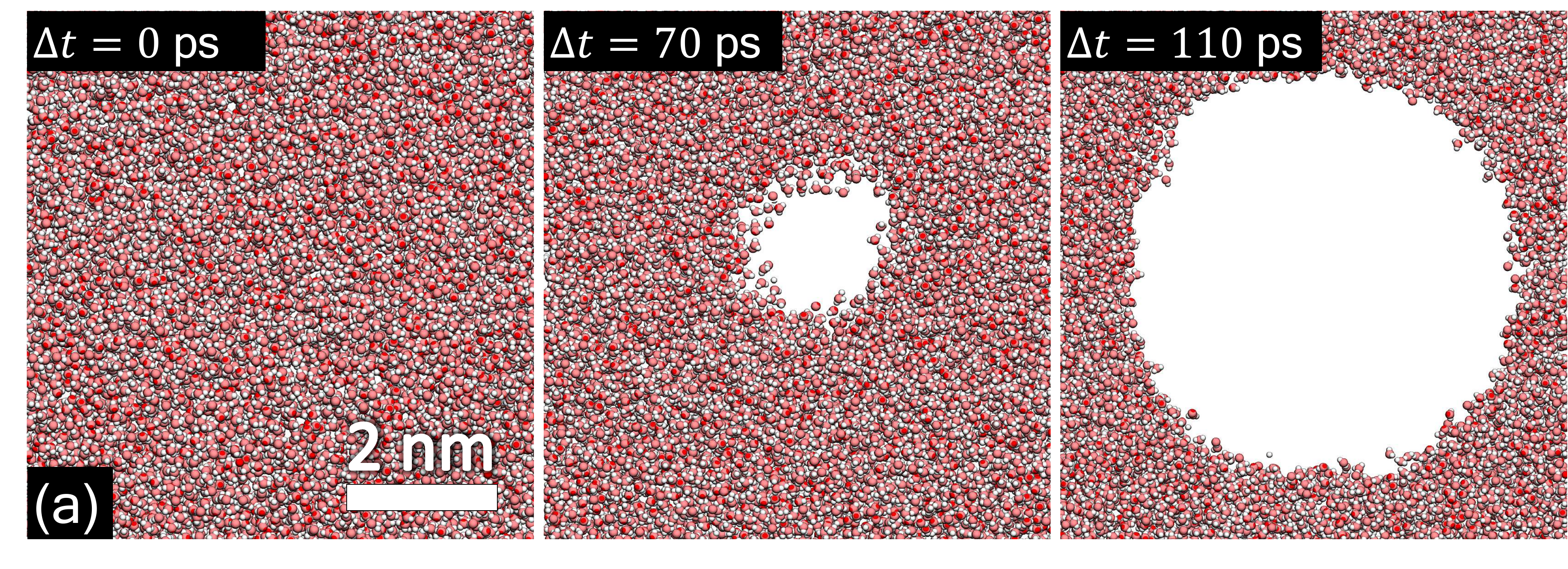}
    % \caption{}
    \end{subfigure}

    \begin{subfigure}[b]{0.45\textwidth}
    \includegraphics[width=\textwidth]{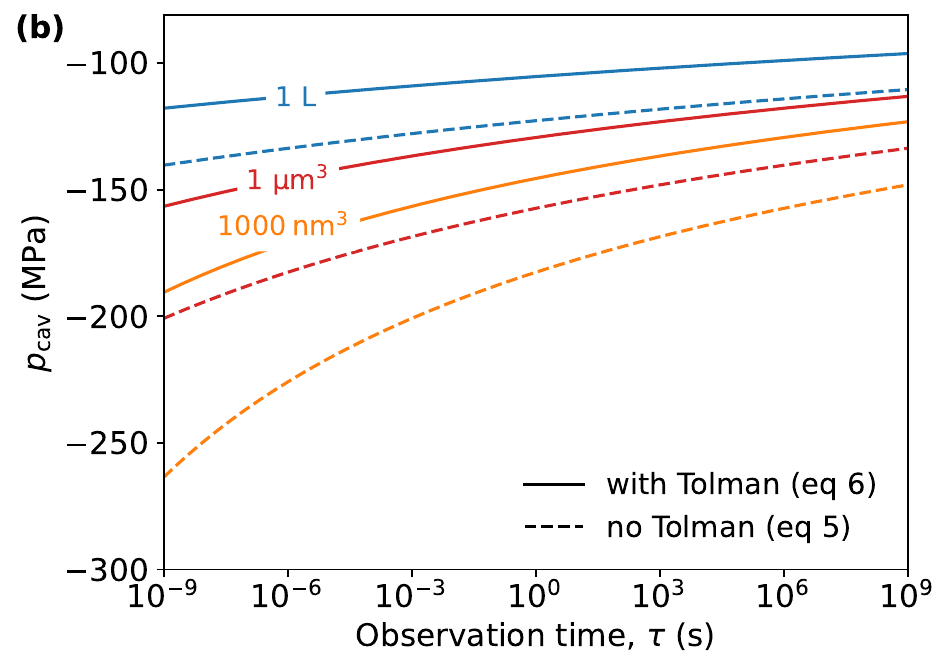}
    % \caption{}
    \end{subfigure}
\caption{
\comment{(a) Sequential snapshots of a cavitation event in absolutely pure bulk water from a pressure-ramp simulation, occurring as the ramping pressure reached $p_\mathrm{cav}^* = -178$ MPa under an imposed pressure rate of $\dot{p} = -$1 MPa ns$^{-1}$. Each snapshot shows a thin cross-sectional slab through the cavitation site. The values of $\Delta t$ indicate the elapsed time from a reference point just at cavitation onset.}
(b) Cavitation pressure as a function of observation time for three different volumes of bulk TIP4P/2005 water. Dashed lines show the result without the Tolman correction (\Eq~\ref{eq:tens_strength_old}), and the full lines are the results with the Tolman correction (\Eq~\ref{eq:p_cav_vs_t}).
}
\label{fig:p_vs_t_cav}
\end{figure}

\comment{When simulating cavitation, it is essential that the model accurately reproduces surface tensions. After evaluating four popular water models, we selected TIP4P/2005~\cite{tip4p_water}, which has a surface tension of 68 mN/m, closely matching the experimental value.} 
In our simulations, we employ the pressure-ramp technique~\cite{boucher2007pore, Kanduc10733}, where the negative pressure decreases linearly over time, as outlined in the Methods section. In this approach, cavitation \comment{(an example of which is shown in \Fig~\ref{fig:p_vs_t_cav}a)} occurs at a dynamic cavitation pressure dependent on the applied pressure rate.
From the relationship between dynamic cavitation pressures and pressure rate \comment{(\Fig~\ref{fig:fits}a in the Methods section)}, we derive two key parameters:
the Tolman length ($\delta_\trm{w}=-0.058$ nm) and the attempt frequency density ($\kappa_\trm{w}=1.25 \times 10^{18}$ s$^{-1}$nm$^{-3}$).

% \begin{figure}[h]\begin{center}
% \begin{minipage}{0.45\textwidth}\begin{center}
% \includegraphics[width=\textwidth]{figures/p_cav_vs_t_h2o.pdf} 
% \end{center}\end{minipage}
% \caption{
% Cavitation pressure as a function of time for three different volumes of bulk TIP4P/2005 water. Dashed lines show the result without the Tolman correction (\Eq~\ref{eq:tens_strength_old}) and the full lines are the results with the Tolman correction (\Eq~\ref{eq:p_cav_vs_t}).
% }
% \label{fig:p_vs_t_cav}
% \end{center}\end{figure}

These parameters were then used in the CNT model (\ref{eq:tens_strength_old} and \Eq~\ref{eq:p_cav_vs_t}) to calculate the cavitation pressures for variable volumes at static conditions. Figure \ref{fig:p_vs_t_cav}b shows these pressures for three different  volumes of absolutely pure water against observation time (solid lines). Longer observation times and larger volumes result in slightly less negative cavitation pressures. However, their influence is very weak, as they combine within the logarithmic function under the square root in \Eq~\ref{eq:tens_strength_old}. For a wide range of practical scenarios, ranging from microseconds to years ($\sim10^8$~s) and involving water volumes from cubic micrometers to liters, cavitation pressures span between $-100$ and $-150$ MPa. This relatively narrow window allows us to consider the cavitation pressure, in a rough approximation, as the tensile strength of water in the sense of an intrinsic material property. 
Additionally, for typical simulation volumes of $10\times10\times10$~nm$^3$ (solid orange line), the cavitation pressures are around $-190$ MPa, close to the dynamic cavitation pressures in our pressure-ramp simulations (\Fig~\ref{fig:fits}a in the Methods section).

The negative Tolman length in our water model reduces the critical bubble radius (\Eq~\ref{eq:r_star_tolam}) and lowers the free energy barrier (\Eq~\ref{eq:dG_star_approx}), which facilitates cavitation. 
 For comparison, we also show predictions without the Tolman correction (\Eq~\ref{eq:tens_strength_old}), represented by dashed lines in \Fig~\ref{fig:p_vs_t_cav}b, consistent with other theoretical studies~\cite{fisher1948fracture, caupin2005liquid, caupin2006cavitation, caupin2013stability, azouzi2013coherent}.
These values are about 20\% higher in magnitude than those with the correction.
Although often disregarded in CNT models to avoid complexity, the Tolman correction significantly impacts results, highlighting the importance of accounting for curvature-dependent surface tension for accurate quantitative predictions.

It is important to recognize that the outcomes of simulations are influenced by the choice of the water model. For reference, we conducted additional simulations using the OPC water model~\cite{opc_water}, known for its very accurate reproduction of the surface tension of actual water, yielding $\gamma_\w=72$ mN/m. As expected, the OPC model results in slightly more negative cavitation pressure values. However, the difference is relatively minor, amounting to only a few percent. This small variance underscores the robustness of the conclusion that ideally clean water cavitates at pressures between $-100$ and $-150$ MPa, in agreement with other theoretical predictions~\cite{fisher1948fracture, caupin2005liquid, caupin2006cavitation, caupin2013stability, azouzi2013coherent}.

Yet, a persistent question is why most carefully conducted experiments with ultra-purified water (purified and degassed as much as possible) consistently report cavitation pressures that are 4 to 5 times lower than the theoretical value~\cite{caupin2006cavitation, caupin2013stability, caupin2015escaping}. The harsh reality is that achieving absolute purity in liquids is virtually impossible, \comment{as trace impurities are almost always present~\cite{roger2012hydrophobic, maali2017viscoelastic}.} Notably, even the highest grade of ultrapure water achievable with current technology still contains total organic carbon (TOC) levels above 1 \si{\ug}/L~\cite{uematsu2019impurity, zhang2021critical}. In a cubic millimeter of such water, the entire organic matter is equivalent in volume to a droplet of about 1 \si{\um} in diameter. While smaller and medium-sized organic molecules may dissolve, 
\comment{larger molecules, such as typical phospholipids~\cite{serravalle2024critical} or alkanes with more than 12 carbon atoms~\cite{letinski2016water} have solubilities comparable or below this TOC level and thus tend to stay aggregated.} Consequently, hydrocarbon aggregates are inevitably present in any macroscopic volume of water, even when it is highly purified~\cite{berkelaar2014exposing}. This notion brings us to an important question: could these tiny aggregates---referred to here as droplets---even in minimal quantities, impact the tensile strength of the entire aqueous system? 
Addressing this question is the central aim of the remaining part of our study.

\subsection{Homogeneous cavitation in bulk liquid hydrocarbon}
Before delving into the impact of small hydrophobic droplets, we first analyze homogeneous cavitation in a pure liquid hydrocarbon subjected to negative pressure. This setup effectively represents cavitation inside a large oil droplet, as the expansion of the outer droplet boundary becomes negligible before the nucleating bubble inside the droplet reaches the critical size. Consequently, the expressions for the cavitation pressure in this scenario are equivalent to that of bulk water (\Eqs~\ref{eq:tens_strength_old} and \ref{eq:p_cav_vs_t}) with appropriate substitutions of the quantities to account for the properties of the oil, specifically, $\gamma_\trm{w}\to\gamma_\trm{o}$, $\delta_\trm{w}\to\delta_\trm{o}$, and $\kappa_\trm{w}\to\kappa_\trm{o}$.
To derive these parameters, we extended our simulation to include bulk decane, conducting them in an analogous way as performed with bulk water. 
\comment{We employed the CHARMM36 force field for decane, which accurately reproduces decane--vapor~\cite{10.1093/pnasnexus/pgad190} and decane--water surface tensions.}
\comment{Pressure-ramp simulations of bulk decane (results shown in \Fig~\ref{fig:fits}b in the Methods section) provided the values for $\kappa_\o$ and $\delta_\o$ listed in Table~\ref{tab:parameters}.}

The cavitation pressures for bulk decane based on \Eqs~\ref{eq:tens_strength_old} and \ref{eq:p_cav_vs_t} behave qualitatively similar to those for water, but fall within the range of $-20$ to $-30$ MPa for volumes ranging from cubic micrometers to liters and observation times from microseconds to days.
\comment{This range is consistent with measured cavitation pressures in many other organic liquids~\cite{galloway1954experimental, apfel1977tensile, ohde1993raising, vinogradov2000boundary}.
The reduced threshold in hydrocarbons arises primarily from their lower surface tension values, $\gamma_\o\approx20$--$30$ mN/m, which are comparable across most organic liquids~\cite{bormashenko2010values} and significantly lower than that of water.} The surface tension of hydrocarbons is primarily due to weak cohesive forces driven by dispersion (London) interactions, whereas the high surface tension of water and corresponding strong cohesion are due to hydrogen bonding.
Overall, the cavitation pressure in hydrocarbons is roughly five times lower in magnitude than that of bulk water. 
This difference can be expressed analytically from \Eq~\ref{eq:tens_strength_old} for water and oil parameters in the macroscopic limit, where $Vt\to\infty$, and one indeed finds ${p}^{}_{\trm{cav,o}}/{p}^{}_{\trm{cav,w}}\approx(\gamma_\o/\gamma_\w)^{3/2}\approx 1/5$.
Thus, the disparity in tensile strengths between water and hydrocarbon is primarily influenced by the different surface tensions of the two liquids. 

%\MK{comment on the Tolman length correction?}

\subsection{\comment{Cavitation of water containing oil nanodroplets}}

%We will assume that the water volume is much larger than the volume of the droplet.
\begin{figure*}[t]\begin{center}
\begin{minipage}[b]{0.95\textwidth}\begin{center}
\includegraphics[width=\textwidth]{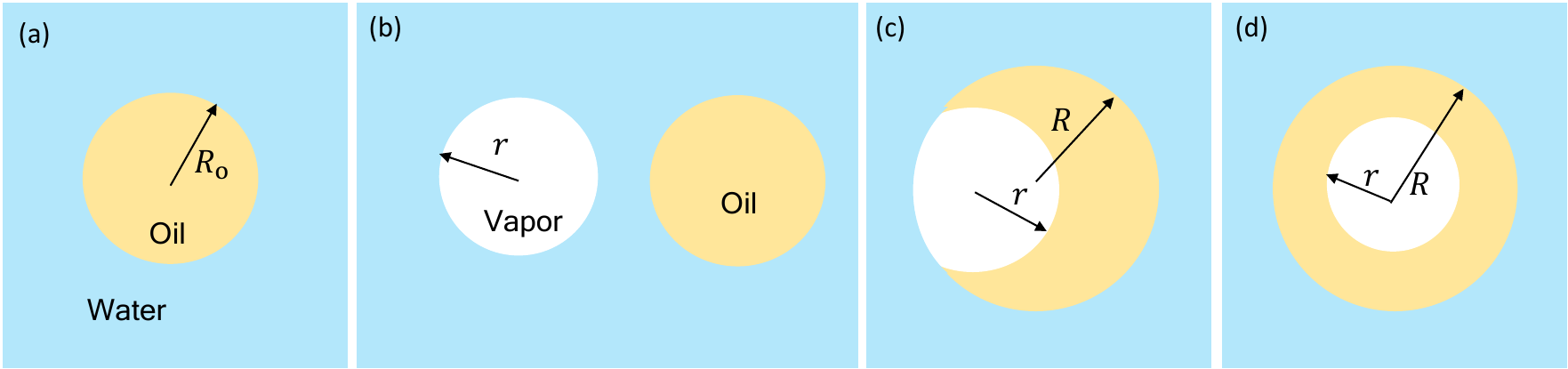}
\end{center}\end{minipage}
\caption{
(a) Oil droplet of initial radius $R_\o$ in water can undergo three cavitation scenarios:
(b) homogeneous cavitation inside the bulk water phase, (c) heterogeneous cavitation at the droplet interface, and (d) cavitation inside the oil droplet.
}
\label{fig:droplet_cavitation}
\end{center}\end{figure*}
An intriguing question is whether nonsoluble hydrocarbons also influence cavitation when present as nanoscopic droplets in water. To investigate this, we proceed by considering a macroscopically large volume of water, $V_\trm{w}$, containing a single oil droplet of radius $R_\o$, as illustrated in \Fig~\ref{fig:droplet_cavitation}a. \comment{We then extend our analysis to a system with $ N$ monodisperse oil droplets.}

Under negative pressures, this system can undergo three generic cavitation scenarios: The bubble can nucleate (i) in the bulk water phase (\Fig~\ref{fig:droplet_cavitation}b), (ii) at the droplet--water interface (\Fig~\ref{fig:droplet_cavitation}c), or (iii) inside the oil droplet (\Fig~\ref{fig:droplet_cavitation}d).

These three scenarios represent three distinct pathways for the cavitation process within the system. Each pathway evolves with its specific rate, and the overall cavitation rate of the system is the sum of all the contributions, assuming they proceed independently.
\comment{For generality, we express the overall cavitation rate for a system with $N$ monodisperse oil droplets as}
\begin{equation}
k = k_{\text{w}} + \comment{N}k_{\text{ow}} + \comment{N}k_{\text{o}}
\label{eq:cav_rate}
\end{equation}
The first pathway, homogeneous cavitation within the bulk water phase, is characterized by the cavitation rate $k_\trm{w}$ and is given by \Eq~\ref{eq:k}.
Moving on to cavitation in the droplet's interior (\Fig~\ref{fig:droplet_cavitation}d), the cavitation rate for a single droplet can be expressed as
\begin{equation}
    \label{eq:cav_rate_hom}
    k_\o = \kappa_\o V_\o e^{-\beta G^*_\o}
\end{equation}
where $\kappa_\o$ is the attempt frequency per volume of oil, derived from our bulk decane simulations in the previous section, $V_\o = ({4 \pi}/{3}) R_\o^3$ is the volume of the oil droplet, and $G^*_\o $ is the corresponding free energy barrier.
The latter is the maximum in the free energy of a bubble nucleating within the droplet, which we formulate  as
\begin{eqnarray}
    \label{eq:DeltaGo}
    G_\o(r) &=& 
    4 \pi \gamma_\o (1+2\delta_\o/r)r^2 \nonumber\\
    &&+ \>4 \pi \gamma_\ow \left[R(r)^2 -R_\o^2\right] +  \frac 43 \pi p r^3 
\end{eqnarray}
and is associated with the expansion of two interfaces: the inner one of radius $r$, characterized by the oil--vapor surface tension $\gamma_\o$, and the outer interface of radius $R$, characterized by the oil--water surface tension $\gamma_\ow$. The two radii are linked through the reasonable assumption that the oil volume remains constant, leading to $R^3=R_\o^3+r^3$. 
In the above expression, we have only included the term associated with the curvature correction of the inner oil--vapor interface (described by the Tolman length $\gamma_\o$). We have omitted the curvature correction of the outer oil--water interface, reading $8 \pi \gamma_\ow \delta_\ow (R-R_\o)$, where $\delta_\ow$ is the Tolman length of the oil--water interface, which we have not quantified. 
This omission is justifiable, especially for larger droplets where $R_\o$ significantly exceeds $r$. In such cases, we can approximate $R-R_\o\approx r^3/(3 R_\o^2)$, 
resulting in a much smaller contribution from the curvature correction of the oil--water interface compared to the curvature correction of the internal, oil--vapor interface.
%This simplification allows us to fully evaluate the cavitation rate in the droplet's interior (\Eq~\ref{eq:cav_rate_hom}) based on the parameters form the bulk decane simulations.
The maximum in $G_\o(r)$ does not lend itself to a simple analytical expression and  requires numerical evaluation.

%%%%%%%%%%%%%%%

Finally, the rate of heterogeneous cavitation at the oil--water interface of a single droplet (\Fig~\ref{fig:droplet_cavitation}c) is governed by the equation
\begin{equation}
    \label{eq:cav_rate_het}
    k_\ow = \kappa_\ow A_\o e^{-\beta G^*_\ow}
\end{equation}
where $\kappa_\ow$ represents the cavitation attempt frequency per surface area of the droplet, $A_\o = 4 \pi R_\o^2$ is the droplet's surface area, and $G^*_\ow$ is the associated free energy barrier.

\begin{figure}[h]
\begin{minipage}[b]{0.45\textwidth}\begin{center}
\includegraphics[width=\textwidth]{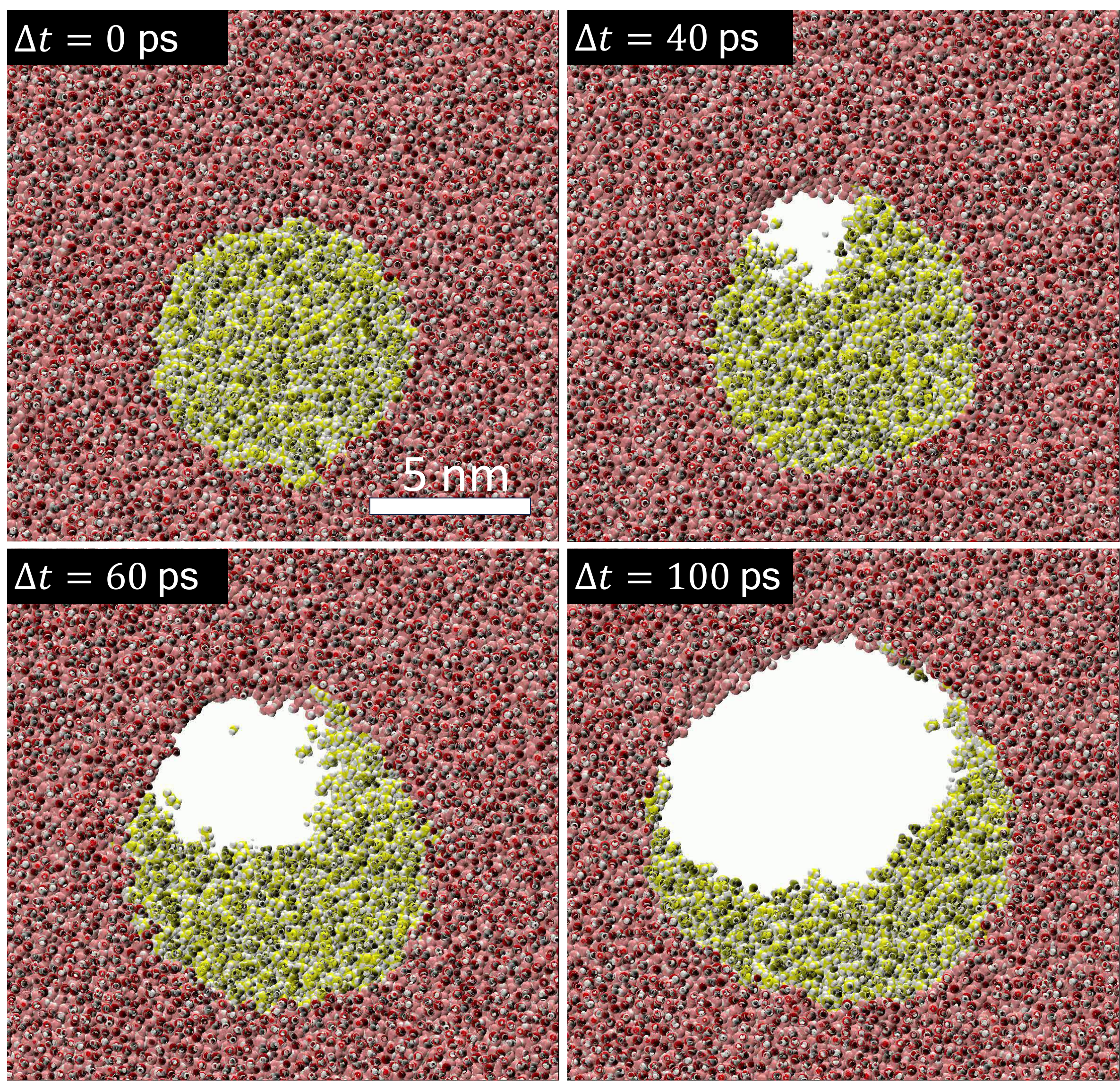}
\end{center}\end{minipage}
\caption{
Sequential snapshots from a pressure-ramp simulation at a rate of $\dot{p}=-1$ MPa ns$^{-1}$, showing a cavitation event of a decane droplet with an initial radius of $R_\o = 4.2$~nm in water.
The cavitation occurred as the pressure decreased to $p_\mathrm{cav}^* = -63$ MPa.
Each snapshot represents a thin cross-sectional slab through the droplet center.
%\comment{Sequential snapshots from a pressure-ramp simulation illustrating a cavitation event within a decane droplet in water. The snapshots depict thin cross-section slabs through the center of the droplet. The times $\Delta t$ show the time passed from a reference point before the onset of cavitation. The radius of the decane droplet before cavitation occured is $R_\mathrm{o}=4.2$ nm. This cavitation event occurred when the pressure decreased down to $p_\mathrm{cav}^*=-$63 MPa.}
}
\label{fig:droplet_sim_snapshot}
\end{figure}

To quantify this cavitation pathway, we performed a series of simulations with decane droplets with radii ranging from 2.5 to 6.5 nm. 
Density and order-parameter profiles indicated a rapid transition to an isotropic bulk-like region inside the droplet, with an effective water--decane interface being approximately half a nanometer thick.
Exposing these droplets to increasing negative pressures, we consistently found that cavitation begins at the oil--water interface, thus, corresponding to the heterogeneous cavitation scenario.
An example of such cavitation is captured by sequential snapshots in \Fig~\ref{fig:droplet_sim_snapshot} for a droplet with an initial radius of 4.2 nm.
%Prior to conducting numerical analysis, two distinct scenarios can be anticipated. In the case of a small or diminishing droplet, $R_\trm{o}\to 0$, the first term, $k = k_\w$, is expected to predominantly influence the cavitation rate. Conversely, for a very large droplet, $R_\trm{o}\to \infty$, the third term becomes prevalent. This is due to the lower free energy barrier compared to bulk water, and the fact that the attempt frequency scales with volume, which eventually beats the surface area scaling of the second term. A nontrivial question arises regarding the intermediate stages: whether heterogeneous cavitation, as described by the second term, occurs at all, or if it is superseded by the third term as the droplet grows. The answer hinges on specific values of attempt frequencies and the free energy barrier.

Heterogeneous nucleation at the droplet interface involves three phases in direct contact, resulting in a more complex geometry than in the other two pathways. Because the shapes do not conform to equilibrium geometries~\cite{binyaminov2021thermodynamic}, formulating an expression for $G^*_\ow$ becomes challenging.
As seen from the snapshots, the nucleating cavity  expands predominantly into the oil phase, while the water phase retains a nearly spherical interface with both the oil and the vapor. 
Conceptually, this scenario can be envisioned as bringing a spherical bubble from the droplet's interior to its outer interface with water, \comment{creating a water--vapor interface while eliminating the oil-bound interfaces at the patch.
Antonov's rule states that the interfacial tension between two liquids approximately equals the difference in their respective surface tensions with vapor, that is, $\gamma_\trm{o} + \gamma_\trm{ow} \approx \gamma_\trm{w}$~\cite{makkonen2018another}.
 This relation implies that the transition of a spherical bubble results in minimal net change in free energy.
}
%Given the approximate equality $\gamma_\trm{o}+\gamma_\trm{ow}\approx \gamma_\trm{w}$, often referred to as Antonov's rule in the context of hydrocarbon materials~\cite{makkonen2018another}, this process results in minimal net change in the free energy.
%Therefore, in a reasonable approximation, the free energy barriers for heterogeneous and in-droplet cavitations are similar, $\Delta G_\trm{ow}^*\approx \Delta G_\trm{o}^*$.

Thus, the distinction between $G_\trm{ow}$ and $G_\trm{o}$ arises primarily from additional correction terms in the former, such as the Tolman curvature corrections of the water--vapor interface and the contribution of the three-phase contact line where oil, water, and vapor meet.
Quantifying these contributions is challenging.  Since all these factors scale linearly with the bubble radius, $r$, we adopt the expression for the free energy as given by \Eq~\ref{eq:DeltaGo} but substitute the oil--vapor Tolman length with an effective value, thus $G_\trm{ow} = G_\trm{o}(\delta_\o\to\delta_\trm{eff})$. This ``effective Tolman length'' serves as a fitting parameter to our simulation results, offering a pragmatic approach to encompass these complexities. 
\comment{Performing pressure ramp simulations and fitting their results (shown in \Fig~\ref{fig:fits}c in the Methods section) yield the parameters 
$\kappa_\trm{ow}$  and $\delta_\trm{eff}$, listed in Table~\ref{tab:parameters}.}
We note that the global fit is not perfect, likely reflecting the complexity and dynamic nature of the cavitation geometry. A better fit would require higher-order corrections in the free energy or even in the attempt frequency density, which, however, is beyond the scope of this study.

%, thus $\Delta G^*_\ow\approx \Delta G^*_\o$.
 
\begin{figure}[h]
\begin{center}
\begin{minipage}[b]{0.45\textwidth}\begin{center}
\includegraphics[width=\textwidth]{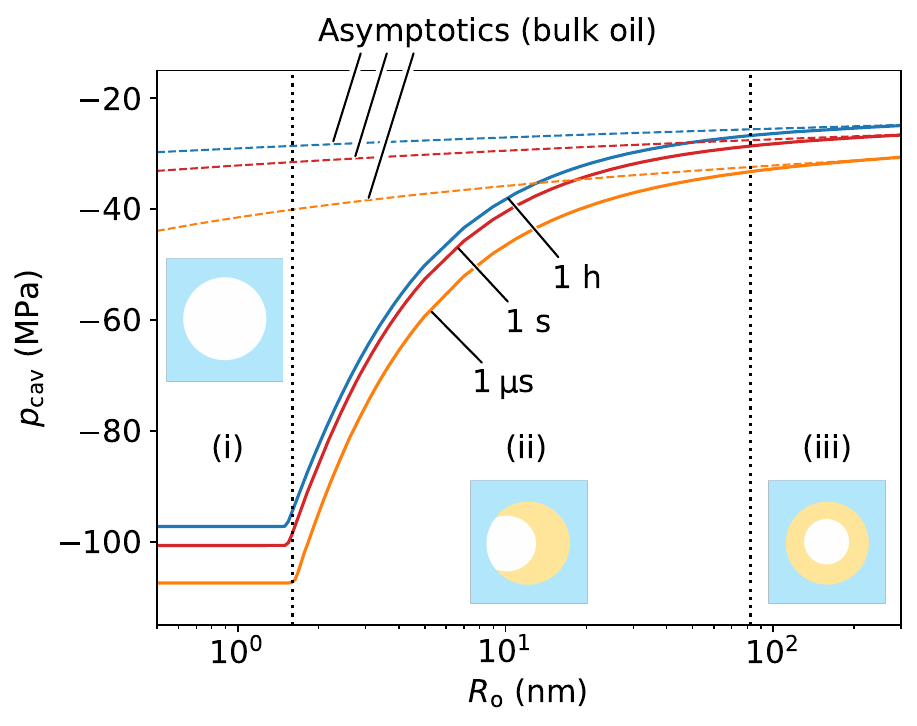}
\end{center}\end{minipage}
\caption{
Cavitation pressure (tensile strength) of 1~liter of water containing a single decane droplet \comment{($N=1$)} plotted versus its radius for various observation times $\tau$ (solid lines). 
With an increasing droplet size, the cavitation starts as homogeneous cavitation in bulk water [region (i)], crosses over at $R_\o\approx 1.5$ nm to heterogeneous cavitation at the droplet surface [region (ii)], and at $R_\o \approx 80$ nm eventually crosses over to cavitation in the droplet's interior [region (iii)].
For large droplets, the cavitation pressures asymptotically approach the results for bulk decane of equivalent volume (shown by dashed lines).
}
\label{fig:tens_strength_dec}
\end{center}\end{figure}

Equipped with all the required model parameters, we are now set to calculate the cavitation pressure by solving \Eq~\ref{eq:cav_rate}. Unlike the scenario of cavitation in bulk liquid, where an analytical expression was possible
(provided by \Eqs~\ref{eq:tens_strength_old} and \ref{eq:p_cav_vs_t}), we have to rely on numerical methods in this case. Doing so, we plot the calculated cavitation pressure for $V_\trm{w}=1$ liter of water \comment{containing a single decane droplet ($N=1$) as a function of its} radius $R_\o$ for three observation times in \Fig~\ref{fig:tens_strength_dec}. This plot represents the central finding of our study.

In the case of a vanishing droplet, $R_\o\lesssim 1$~nm \comment{[region (i)]}, the cavitation rate is dominated by the first term in \Eq~\ref{eq:cav_rate} (homogeneous cavitation in water), and the cavitation pressure of the system converges to the tensile strength of absolutely pure water, as  given by \Eq~\ref{eq:p_cav_vs_t}.
A notable transition occurs when the droplet radius exceeds approximately $1.5$ nm.
At this point, the cavitation pressure abruptly increases to less negative values \comment{[region (ii)]}. This transition occurs when the rate of heterogeneous cavitation reaches that of the bulk water phase, $k_\trm{w}=k_\trm{ow}$ \comment{(denoted by the left vertical dotted line)}. 
The total volume of the system, $V_\trm{w}$, has a minor effect on this transition.

As the droplet radius continues to increase, the predominant cavitation mechanism transitions from heterogeneous to in-droplet cavitation \comment{[region (iii)]}. 
The crossover between the two pathways, i.e., when their rates are equal, i.e., $k_\trm{ow}=k_\trm{o}$,  occurs at a droplet radius of approximately 80 nm \comment{(denoted by the right vertical dotted line)}. 
Beyond this size, bubble nucleation primarily occurs within the hydrocarbon phase of the droplets.

Two key insights emerge from \Fig~\ref{fig:tens_strength_dec}.
The first and most important insight is that remarkably tiny amounts of nonpolar material are sufficient to substantially reduce the cavitation pressure of an entire macroscopic aqueous system.  
A single oil nanodroplet, only a few nanometers in size, weakens the tensile strength from the theoretical value for absolutely pure bulk water (around $-$100 MPa for 1 liter) down to the tensile strength typical of hydrocarbons, which is \comment{around $-$30 MPa~\cite{galloway1954experimental, apfel1977tensile, ohde1993raising}.}
As mentioned above, avoiding such tiny insoluble droplets is basically impossible.
The other insight from \Fig~\ref{fig:tens_strength_dec} is that once the droplet radius exceeds several tens of nanometers, any further increase in its size has a rather modest impact on the further decrease in the tensile strength.

\comment{Increasing the number of oil droplets in the system, $N$, has only a minor effect on the cavitation pressure. When cavitation is dominated by droplets, the rate simplifies to $k/N \approx k_{\text{ow}} + k_{\text{o}}$, indicating that cavitation pressure depends on $N$ in the same weak manner as it does on observation time $\tau$. For cavitation, the critical factor remains the size of the largest oil droplet.}

This observation underscores a challenging aspect of water purification: regardless of the extent of purification, preventing cavitation becomes a matter of eliminating the very last oil droplet in the system. Namely, that very last droplet is the ``weakest link'' and inevitably contributes to cavitation. 
Consequently, we assert that the cavitation pressures observed experimentally in aqueous systems reflect the presence of minute amounts of nonpolar residues, even in the most thoroughly purified samples.

Despite the points discussed above, nanoscopic oil droplets may be avoidable only in very small volumes of ultrapure water, specifically those approaching the micron scale. In a cubic micrometer of ultrapure water of a TOC purity level of 1 \si{\ug}/L, the condensed organic content would amount to an oil droplet with a radius of one nanometer (equivalent to 50 decane molecules).
However, since smaller organic molecules dissolve at this TOC, micrometer-sized water volumes may indeed realistically be free of organic aggregates larger than around 1~nm in size. This absence could elucidate the outcomes of experiments with microsized water inclusions in minerals, which have reported cavitation pressures as high as $-$140 MPa~\cite{zheng1991liquids, alvarenga1993elastic, azouzi2013coherent}, aligning with our theoretical predictions for absolutely pure water of volume $V_\trm{w}=1$~\si{\um}$^3$ (red solid line in Fig.~\ref{fig:p_vs_t_cav}).

%Two important conclusions can be drawn from \Fig~\ref{fig:tens_strength_dec}.  The first and the most important one is the notion of how remarkably small amount of a nonpolar hydrocarbon is enough to reduce the tensile strength of the system from values for theoretically pure bulk water, $-$110 MPa, down to the tensile strength of hydrocarbons, which are several tens of MPa. A single oil nanodroplet a few nanometers across is enough to compromise the tensile strength of the entire macroscopically large aqueous system. It is important to emphasise that having only one oil droplet in 1 liter of otherwise perfectly pure water would be considered ultrapure water by most criteria.\cite{standard2013standard, melnik2019ultrapure} Ultrapure water is the highest level of water purity achievable with current technology, used only for semiconductor production and far more pure compared to other forms of water that are typically considered pure, such as distilled water or water used in medicine for intravenous injections.\cite{standard2013standard, melnik2019ultrapure}... dissolved gases as well as nanoscale contaminants are virtually impossible to eliminate completely from any substantial liquid volume. It is hard to observe cavitation nuclei at the nanoscale, and thus it is challenging to quantify their effects on cavitation through experimental research.

\comment{Finally, while our theoretical approach provides a foundational understanding of cavitation at nanoscopic hydrophobic inclusions, it has certain limitations.
One of the limitations is the computational costs of MD simulations, which limit our ability to simulate much larger systems or extend them to longer time scales. Consequently, we have to rely on the integration with CNT, which has its own simplifications and shortcomings~\cite{erdemir2009nucleation, karthika2016review}. For instance, we have used a simplified analytic expression for $\Delta G_\ow$, potentially contributing to the suboptimal fit in \Fig~\ref{fig:fits}c. 
Another limitation concerns the realism of our model. We assume no dissolved gases within the oil phase, while, in real scenarios, gases and gas pockets, especially in larger hydrophobic droplets, may influence cavitation dynamics and could introduce additional effects not captured here. For example, a recent study has shown that a sudden pressure drop in shockwave experiments can trigger the nucleation of gas bubbles from a locally supersaturated liquid with gas~\cite{pfeiffer2022heterogeneous}.}

\section{Conclusions}
Even the most rigorously purified water samples contain insoluble aggregates and droplets. 
By integrating MD simulations with classical nucleation theory, we have quantitatively evaluated the impact of these tiny organic aggregates on cavitation pressure in water. Remarkably, a single nanoscopic oil droplet, merely a few nanometers across, can induce cavitation at pressure magnitudes substantially lower than those predicted for absolutely pure water. Specifically, we find that cavitation pressures drop to levels similar to those in  bulk hydrocarbons, approximately $-$30 MPa, a value consistent with measurements from highly purified water experiments.

The quest for absolute water purity in any macroscopically significant system remains an elusive goal, suggesting that the theoretical tensile strength of water may be inherently unattainable in experimental contexts. However, an intriguing outlier exists in the form of microsized inclusions, which are so small that they could actually be free from clusters formed by impurities, enabling cavitation pressures as high as $-$140 MPa to be documented~\cite{zheng1991liquids, alvarenga1993elastic, azouzi2013coherent}. 
\comment{Understanding the role of hydrophobic inclusions provides valuable insights into xylem transport in plants, where hydrophobic and amphiphilic biomolecules, such as lipids, are present~\cite{schenk2015nanobubbles}. Although these molecules might initially seem like potential nucleation sites, they appear to be unproblematic at pressures of $-$10 MPa found in plant xylem. We intend to investigate their specific role in future work.
}

Our findings extend beyond the realm of cavitation, shedding light on closely related processes such as the boiling of superheated water, bubble nucleation in supersaturated solutions, and even more general nucleation phenomena. The influence of tiny impurities on these processes highlights a broader principle: the profound impact that seemingly negligible components can have on the behavior of large-scale systems. This principle is of paramount importance, not only for advancing our theoretical understanding but also for improving experimental methodologies and technologies in fluid mechanics, material science, and beyond.

\section{Methods}
\footnotesize

\subsection{Simulation details}
Molecular dynamics simulations were conducted using Gromacs 2022.1~\cite{bekker1993gromacs}, with an integration timestep of 2 fs.
The temperature was maintained at 300 K using the v-rescale thermostat~\cite{v-rescale}, except in cases where we deliberately examined temperature dependence. The pressure was controlled by the C-rescale barostat~\cite{c-rescale} with isotropic coupling.

For simulating decane, we used the all-atom CHARMM36 \cite{vanommeslaeghe2010charmm} force field.
Electrostatic and Lennard-Jones (LJ) interactions were treated using particle mesh Ewald (PME)~\cite{PME1, PME2} methods with a 1.4~nm real-space cutoff.
 Unlike traditional approaches that employ a cutoff for handling LJ interactions, we used PME summation of LJ interactions~\cite{wennberg2013lennard} with a real-space cutoff of 1.4 nm.
This approach accurately reproduces experimental values of decane surface tensions  over a range of temperatures, as tested recently~\cite{10.1093/pnasnexus/pgad190}.
We opted for the TIP4P/2005~\cite{tip4p_water} model for water, a 4-point model renowned for its accurate reproduction of experimental interfacial tensions and its compatibility with the CHARMM36 force field.
As a comparison to the TIP4P/2005 model, we also tested the OPC~\cite{opc_water} water model for bulk water simulations. 
While the OPC model effectively reproduces a comprehensive set of bulk properties, including the surface tension, very well~\cite{opc_water}, it demonstrates less favorable results when combined with the CHARMM36 force field for alkanes.

\subsection{Ramping pressure method}

Evaluating the attempt frequency densities for cavitation ($\kappa_\trm{w}$, $\kappa_\trm{ow}$, and $\kappa_\trm{o}$) via simulations at constant negative pressure by measuring the mean cavitation time turns out to be impractical. The primary difficulty lies in the very strong dependence of the cavitation time on the imposed pressure.
There exists only a very narrow window of negative pressures where cavitation events can be observed within feasible simulation times, without overly reducing the free energy barrier. Predicting this optimal range of pressures in advance is difficult.

\begin{figure*}[t]\begin{center}
\begin{minipage}[b]{0.32\textwidth}\begin{center}
\includegraphics[width=\textwidth]{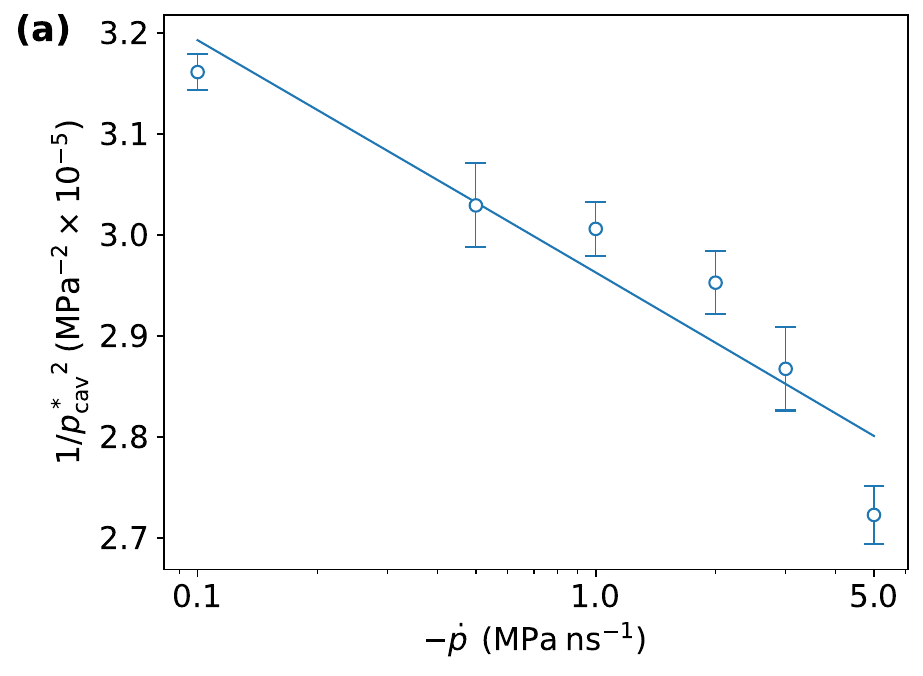}
\end{center}\end{minipage}
\begin{minipage}[b]{0.32\textwidth}\begin{center}
\includegraphics[width=\textwidth]{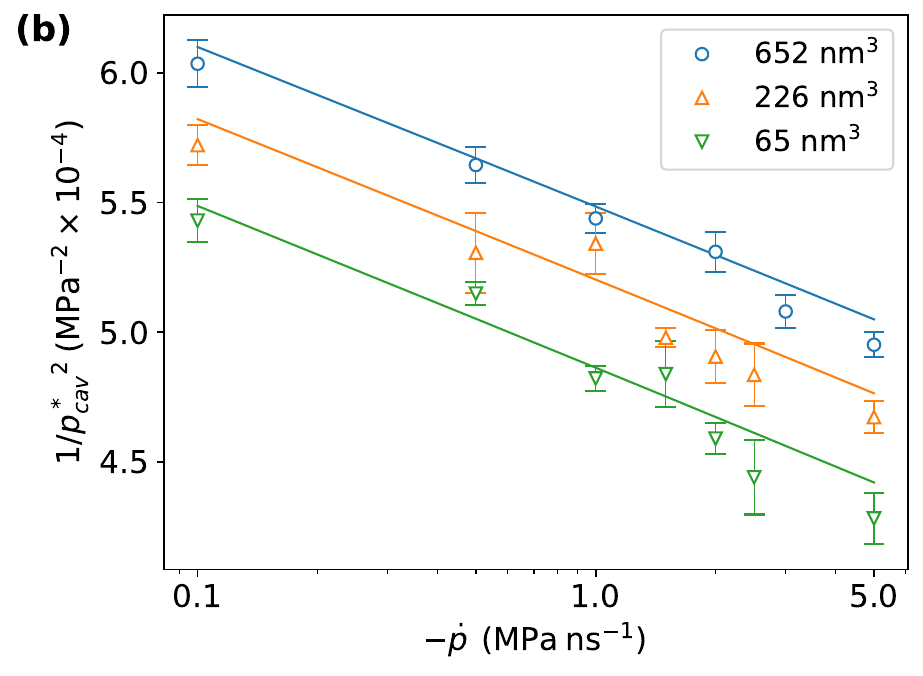}
\end{center}\end{minipage}
\begin{minipage}[b]{0.32\textwidth}\begin{center}
\includegraphics[width=\textwidth]{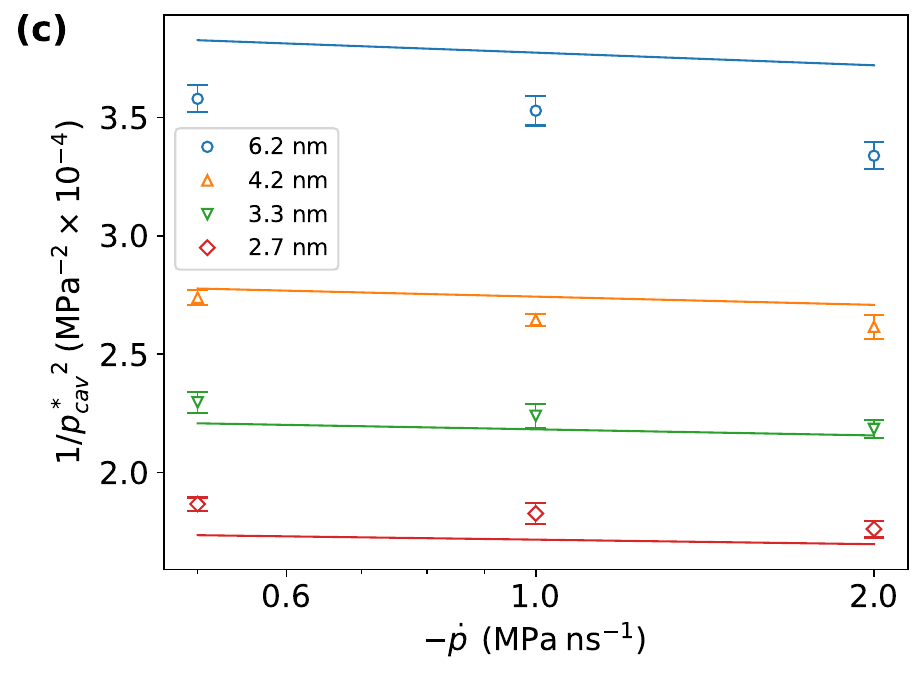}
\end{center}\end{minipage}
\caption{
Mean dynamic cavitation pressures from pressure ramp simulations, plotted as $1/{p^*_\trm{cav}}^2$ against pressure rate for (a) bulk water with a volume of 731 nm$^3$, (b) bulk decane of three different volumes (see legend), and (c) oil droplets with four different initial radii, $R_\o$ (see legend), in water.
Solid lines are fits of \Eq~\ref{eq:p_star} to the data points. In panel (a), the fit corresponds to the single volume, whereas in panels (b) and (c), global fits are simultaneously applied across all three and four datasets, respectively.
%simulations for decane of three different starting volumes. From the fit we obtain $\kappa = 4.09 \times 10^{11} \: \mathrm{s^{-1}nm^{-3}}$ and $\delta = -0.058$ nm. 
%From the fit we obtain $k_0 = 9.13 \times 10^{20}$ s$^{-1}$ and $\delta = -0.029$ nm. Attempt frequency per volume $\kappa = k_0/V = 1.25 \times 10^{18} \: \mathrm{s^{-1}nm^{-3}}$.
%Mean cavitation pressures obtained from constant rate simulations for four distinct droplet sizes in water (with radii ranging from 2.5 to 6.5 nm). From the fit we obtain the attempt frequency per volume $\kappa_\ow = 7.82 \times 10^{13}  \: \mathrm{s^{-1}nm^{-2}}$ and $\delta = -0.076$ nm.
}
\label{fig:fits}
\end{center}\end{figure*}

To circumvent these challenges, we employ the pressure ramp method~\cite{boucher2007pore,
Kanduc10733}, in which we impose a linearly increasing negative pressure over time, defined as $p(t) = \dot p t$, where $\dot{p}$ is the rate of pressure change, assumed to be negative ($\dot{p} < 0$).
Consequently, the free energy barrier decreases over time, reaching a threshold low enough to allow a cavitation event within the simulation time. This protocol 
 renders reaction rates (such as $k_\trm{w}$ in \Eq~\ref{eq:k}) time-dependent. Solving the time-dependent transition rate equations for $k(t)$, as described in Ref.~\citenum{Kanduc10733}, brings us to the following equation
\begin{equation}
    \label{eq:p_star}
    p^*_{\mathrm{cav}} = \dot p \int_0^\infty e^{-k_0 I(t)} \mathrm{d}t
\end{equation}
where $k_0$ is the attempt frequency and
\begin{equation}
    \label{eq:integral}
    I(t) = \int_0^t e^{-\beta G^*(\tau)} \mathrm{d} \tau
\end{equation}
with $G^*(t)$ being the time-dependent free energy barrier as a direct consequence of $p(t)$. Equation \ref{eq:p_star} predicts the mean cavitation pressure, $p^*_{\mathrm{cav}}$, in a system subjected to a pressure ramp at a rate of $\dot{p}$. This ``dynamic'' cavitation pressure $p^*_{\mathrm{cav}}$ should not be confused with the cavitation pressure, $p_{\mathrm{cav}}$, observed under constant pressure conditions, which is our final aim to compute.

Since Gromacs does not support simulating a pressure ramp, we devised a workaround by conducting a sequence of short simulations. Each subsequent simulation was set to a pressure 0.1 MPa lower than its predecessor, effectively emulating a linear pressure decrease. The duration of each simulation depends on the desired pressure rate $\dot p$. We identify a cavitation event based on an abrupt increase in the system's volume after it had linearly risen due to the increasing negative pressure. We employed the random sample consensus (RANSAC)~\cite{ransac} algorithm to accurately pinpoint the moment when the volume begins to increase sharply. 
For each setup, we conducted 8 to 10 independent simulation runs to determine the mean dynamic cavitation pressure, $p^*_{\mathrm{cav}}$, at specific rates of pressure change, 
$\dot p$. In doing so, we plot $1/p_{\mathrm{cav}}^{*2}$ against $-\dot p$, as shown in \Fig~\ref{fig:fits}. Subsequently, we fit \Eq~\ref{eq:p_star} (shown by lines) to the simulation data points with $k_0$ and the Tolman length (which enters the free energy barrier) as fitting parameters.

Specifically, for bulk water (\Fig~\ref{fig:fits}a), the attempt frequency $k_0$ is linked to the simulation volume $V_\trm{w}$ through the relation $k_0=\kappa_\trm{w} V_\trm{w}$ from which $\kappa_\trm{w}$ is extracted.
In the case of bulk decane (\Fig~\ref{fig:fits}b), the attempt frequency is $k_0=\kappa_\trm{o} V_\trm{o}$, where $V_\trm{o}$ is the simulation volume. Given that we conducted simulations over three volumes in this case, we perform a global fit to all three datasets, with $\kappa_\trm{o}$ and $\delta_\trm{o}$ as fitting parameters.
Finally, for heterogeneous cavitation at the droplet interface (\Fig~\ref{fig:fits}c), the attempt frequency is connected to the droplet area as $k_0=4\pi \kappa_\trm{ow} R_\trm{o}^2$.
With four different radii simulated, we again perform a global fit to all four datasets with $\kappa_\trm{ow}$ and $\delta_\trm{eff}$ as the fitting parameters.

\begin{table}[h!]
\footnotesize
    \begin{tabular}{ l | m{3.5cm} } 
        System & Parameter values \\ 
        \hline
        Cavitation in bulk water (w) & 
        $\kappa_\w=1.25 \times 10^{18}$ s$^{-1}$nm$^{-3}$ 
                     $\delta_\w = -0.058$ nm \\ 
        \hline
        Cavitation in bulk decane (o) & $\kappa_\o=5.4 \times 10^{9}$ s$^{-1}$nm$^{-3}$
                      $\delta_\o = -0.116$ nm \\ 
        \hline
        Cavitation at o/w interface & 
        $\kappa_\ow=7.8 \times 10^{13}$ s$^{-1}$nm$^{-2}$ 
                          $\delta_\trm{eff} = -0.076$ nm \\
        \hline
        Water--vapor & $\gamma_\trm{w}=68$ mNm$^{-1}$\\
        Decane--water & $\gamma_\trm{ow}=57$ mNm$^{-1}$\\
        Decane--vapor &$\gamma_\trm{o}=23$ mNm$^{-1}$\\
    \end{tabular}
    \caption{Model parameters obtained from pressure ramp simulations that are used in CNT.}
    \label{tab:parameters}
\end{table}

\subsection*{Acknowledgments}
We thank Fabio Staniscia for useful discussions.
M.\v{S}.\ and M.K.\ acknowledge financial support from the Slovenian Research and Innovation Agency ARIS (contracts P1-0055 and J1-4382).

\comment{
\subsection*{Author contributions}
\textbf{Marin \v{S}ako:} Investigation, Formal analysis, Software, Writing - Original Draft, Writing - Review \& Editing, Visualization. 
\textbf{Roland R. Netz:} Conceptualization, Methodology, Writing - Review \& Editing.
\textbf{Matej Kandu\v{c}:} Conceptualization, Methodology, Writing - Review \& Editing, Supervision, Project administration, Funding acquisition.
}

\comment{
\subsection*{Data availability}
Simulation files and Python scripts used to perform pressure-ramp simulations are available at \url{https://github.com/sajeta/pressure-ramp-md/tree/main}.
}

\normalsize

\footnotesize
\setlength{\bibsep}{0pt}

\bibliography{bibliography/sample}

\end{document}